# Generalized Poincaré Orthogonality: A New Approach to POLSAR Data Analysis

Shane R. Cloude
*AEL Consultants,*
Cupar, Fife, Scotland, UK
e-mail: aelc@mac.com

Ashlin Richardson
*BC Wildfire Service*
Victoria, BC, Canada
e-mail: Ashlin.Richardson@gov.bc.ca

*Abstract*—In this paper we outline a new approach to the analysis of polarimetric synthetic aperture (POLSAR) data. Here we exploit target orthogonality as a multi-dimensional extension of wave orthogonality, familiar on the Poincaré sphere. We first show how to formulate a general basis for a complex orthogonal scattering space using a generalization of the Poincaré formulation, and then show how to optimize the backscattered signal in this space for both monostatic and bistatic radar systems. We illustrate application of the new approach, first to ship detection, using data collected off the north-west of Scotland and then land-use applications in a mixed scene around Glasgow, Scotland, both using L-band ALOS-2 POLSAR data.

*Keywords—synthetic aperture radar, radar polarimetry, orthogonality, image processing.*

## I. Introduction

Polarimetric synthetic aperture radar (POLSAR) is an important radar imaging technology, which enables space-borne, high-resolution imaging of the full 2x2 complex scattering matrix S for each pixel in a scene [1]. Classically this data is then speckle filtered to obtain the average covariance C or coherency matrix T for each pixel, and this is then used as multi-channel input for improved land-use classification or bio-geophysical parameter estimation [2]. For monostatic systems, the reciprocity theorem forces $S = S^T$, and so T and C are 3x3 Hermitian, while for bistatic systems C and T are 4 x 4. Hence there are, per pixel, up to 9 channels for monostatic and 16 for bistatic polarimetry, providing a rich multi-dimensional feature space.

There is however a second key property of polarized waves, so far not widely exploited in SAR, namely orthogonality, whereby we can selectively null a multi-dimensional signal by using a set of complex weights to cancel unwanted contributions and enhance desired features. Potential applications include ship detection at sea, where rough surface scattering can sometimes mask targets, but also land applications, with the ability to isolate and separate land-use classes for further study and analysis.

In this paper we develop a general method for exploiting orthogonality for both monostatic and bistatic POLSAR data. We begin on familiar ground, quickly reviewing the idea of wave orthogonality, familiar via the classical geometry of the Poincaré sphere. We then show how to generalize these concepts to scattering matrices, where a rich space of opportunities arises. We then solve analytically an important problem in this space, namely, how to find the maximum signal in a constrained null sub-space for both bi and monostatic systems. This enables us to find features hidden by an unwanted background. We finally demonstrate application of these ideas using L-band satellite POLSAR data from the ALOS-2 system operated by the Japanese Space agency JAXA.

## II. Orthogonality in Radar Polarimetry

Figure 1 shows a general elliptical polarization state P and its geometrical representation on the surface of the Poincaré sphere. Key for us is that for every state P there is a unique orthogonal partner lying at the antipodal point of this sphere (shown by the red line in fig. 1). Also key, is that while the latitude and longitude of the point P are just the geometrical parameters; orientation θ and ellipticity τ of the ellipse, there exists an alternative angular system, α,δ forming a spherical triangle, as shown in green in fig. 1. It is these latter angles, related to the complex polarization ratio that allow us to extend the idea of orthogonality to higher dimensions.

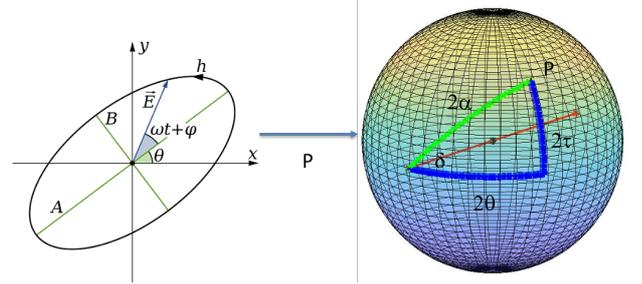

Fig. 1. : Geometry of the Poincaré sphere showing antipodal orthogonal states in red and spherical triangle formed with general point P

To see this, we show in (1) the wave state P as a 2-dimensional vector of electric field components E, in terms of α and δ. We also show the corresponding parameters of the orthogonal state, defined uniquely from the condition of orthogonality, also shown in (1) [1].

$$\underline{E_1} = \begin{bmatrix} E_x \\ E_y \end{bmatrix} = a \begin{bmatrix} cos\alpha \\ sin\alpha e^{i\delta} \end{bmatrix} \xrightarrow{\underline{E_1}^{*T}\underline{E_2}=0} \underline{E_2} = \begin{bmatrix} -sin\alpha e^{-i\delta} \\ cos\alpha \end{bmatrix} \quad (1)$$

We can however extend this idea to higher dimensions by considering the Pauli scattering vector derived from the elements of a monostatic S, as shown in (2) [1,2]. Note that this α angle, although a natural mathematical extension of that in (1), has a different physical interpretation as a ratio of scattering coefficients, not field components [1]. Here we see 5 parameters, with 4 angles defining a unitary vector in a 3-dimensional complex space.

$$\underline{k} = \frac{1}{\sqrt{2}} \begin{bmatrix} S_{HH} + S_{VV} \\ S_{HH} - S_{VV} \\ 2S_{HV} \end{bmatrix} = a \begin{bmatrix} cos\alpha \\ sin\alpha cos\beta e^{i\varphi_1} \\ sin\alpha sin\beta e^{i\varphi_2} \end{bmatrix} = a\underline{e_1} \quad (2)$$

Given any such vector, analogous to the state P in fig.1, we can now define a 2-dimensional orthogonal space (in general for an N dimensional vector we obtain an N-1 dimensional orthogonal space), as shown in (3). The two vectors in (3) span the 2-D null space, but we can form an infinite number of other valid vectors, $\underline{w}$, from linear combinations of these two, as shown in (4).

$$\underline{e_2} = \begin{bmatrix} -sin\alpha \\ cos\alpha cos\beta e^{i\varphi_1} \\ cos\alpha sin\beta e^{i\varphi_2} \end{bmatrix}, \underline{e_3} = \begin{bmatrix} 0 \\ -sin\beta e^{i\varphi_1} \\ cos\beta e^{i\varphi_2} \end{bmatrix} \quad (3)$$

$$\underline{w} = cos\gamma \underline{e_2} + sin\gamma e^{i\tau}\underline{e_3} \rightarrow \begin{cases} \underline{w}^{*T}\underline{w} = 1 \\ \underline{w}^{*T}\underline{e_1} = 0 \end{cases} \quad (4)$$

This process has an underlying geometry analogous to the point P on the Poincaré sphere of fig. 1. In (4) we have parameterized $\underline{w}$, a general unit vector in this space, in terms of 2 angles, similar to $\alpha,\delta$ in (1) (now $\gamma$ and $\tau$). Here we seek ways to exploit the new freedom involved in varying the 2 angles $\gamma$ and $\tau$. To illustrate, we first consider a well-known example, that mirror reflections can be nulled using circular polarization, and we show that this is but a special case of a much wider set of possibilities for analysis.

*A. Mirror cancellation and compact polarimetry*

Mirror reflection has a simple S matrix in radar backscatter, namely the 2x2 identity ($\alpha = 0$ in (2)). Hence the Pauli vector and orthogonal space are simply defined as shown in (5).

$$S_{mirror} = \begin{bmatrix} 1 & 0 \\ 0 & 1 \end{bmatrix} \rightarrow \underline{e_1} = \begin{bmatrix} 1 \\ 0 \\ 0 \end{bmatrix} \rightarrow \underline{e_2} = \begin{bmatrix} 0 \\ 1 \\ 0 \end{bmatrix} \underline{e_3} = \begin{bmatrix} 0 \\ 0 \\ 1 \end{bmatrix} \quad (5)$$

Substituting in (4), we obtain the general null vector as shown in (6).

$$\underline{w} = cos\gamma \underline{e_2} + sin\gamma e^{i\tau}\underline{e_3} = \begin{bmatrix} 0 \\ cos\gamma \\ sin\gamma e^{i\tau} \end{bmatrix} \quad (6)$$

This corresponds to a family of S matrices of the form shown in (7).

$$S_{null} = a \begin{bmatrix} cos\gamma & sin\gamma e^{i\tau} \\ sin\gamma e^{i\tau} & -cos\gamma \end{bmatrix} \quad (7)$$

This family of orthogonal choices is usually hidden, because they all have one property in common, namely they reflect circular polarization like a switch, with 1 or 0 for co or cross-polar, as shown in (8). This is the well-known result that like-circular channels (LL or RR) reduce mirror type reflections, while opposite circular (LR or RL) maximize the return.

$$\begin{cases} \frac{1}{2}[1 \quad i]S_{null}(\gamma,\tau)\begin{bmatrix} 1 \\ i \end{bmatrix} = 1 \rightarrow S_{mirror} = 0 \\ \frac{1}{2}[1 \quad -i]S_{null}(\gamma,\tau)\begin{bmatrix} 1 \\ i \end{bmatrix} = 0 \rightarrow S_{mirror} = 1 \end{cases} \quad (8)$$

This can be demonstrated for example in compact SAR polarimetry [3], which uses a circular transmitter and receives the full Stokes vector of the scattered wave. From these Stokes vectors we can then simulate co and cross circular channels and form images. An example is shown for a pair of images in fig. 2.

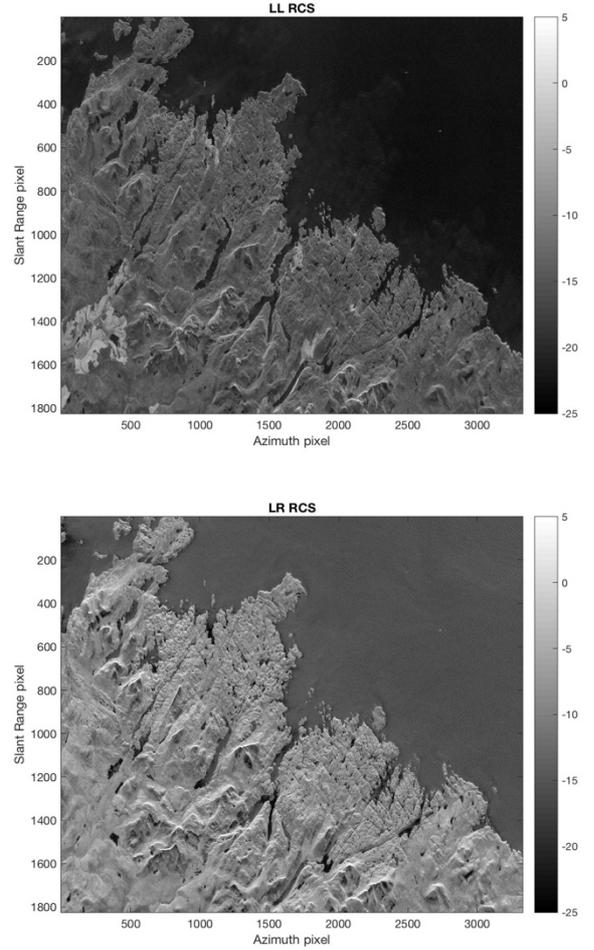

Fig. 2. Co (upper) and cross (lower) circularly polarized imagery, simulated from ALOS-2 L-Band quadpol data for NW of Scotland.

This is simulated compact data from ALOS-2 POLSAR data for a scene off the north west coast of Scotland (collected 8th March 2016). We see the upper copolarized circular return (mirror cancellation) leads to much reduced sea surface scattering (and ships can be better detected), while the lower cross-polarized LR signal (mirror enhancement) emphasises wave features in the ocean. Note that land features also change in the two channels. This illustrates how we can use orthogonality to change visible structure in SAR images, but is only a very special (symmetric) case of the more general cancellation opportunities implicit in (4). In general, no single polarization state (like circular) will provide the on/off access of the mirror for general scattering. Instead, we must employ a more general approach based on measurement of the full S matrix itself. We now turn to consider the general case, first for monostatic then for bistatic systems.

III. THE GENERAL MONOSTATIC CANCELLER

For every general vector $\underline{w}$ in the orthogonal space (see (4)), we can define its orthogonal partner, as shown in (9). This leads to a geometrical interpretation of the monostatic null space as points on a sphere, a generalization of the Poincaré sphere, as shown in fig. 3. This we call an ortho-sphere.

$$\begin{cases} \underline{w} = cos\gamma \underline{e_2} + sin\gamma e^{i\tau}\underline{e_3} \\ \underline{w_\perp} = cos\gamma \underline{e_3} - sin\gamma e^{-i\tau}\underline{e_2} \end{cases} \quad (9)$$

Here again we see an antipodal orthogonal basis $\underline{e_2},\underline{e_3}$ (now complex 3 vectors) and the use of 2 spherical angles $\tau$ and $\gamma$ to locate an arbitrary $\underline{w}$ vector in the null space. This then has its own orthogonal partner at the unique antipodal point.

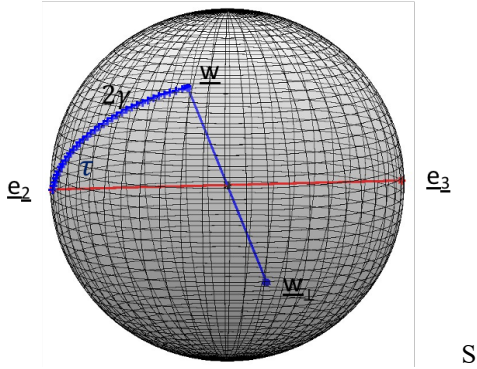

Fig. 3. The ortho-sphere, showing a base pair $e_2,e_3$ that define the orthogonal space, and how to employ 2 spherical angles to locate an arbitrary vector $\underline{w}$, with its own antipodal orthogonal partner.

However, as we move over the surface of this sphere, the projection of $\underline{w}$ onto any given Pauli vector $\underline{k}$ will change. It is then of interest to find the maximum signal in this space (represented by the surface of the sphere in fig.3). To see this, we start with a general S matrix, with corresponding Pauli vector $\underline{k}$ as shown in (10) [1].

$$\underline{k} = \frac{1}{\sqrt{2}}\begin{bmatrix} S_{HH} + S_{VV} \\ S_{HH} - S_{VV} \\ 2S_{HV} \end{bmatrix} \rightarrow m = \underline{w}^{*T}\underline{k} \rightarrow D(\underline{w}) = mm^* \quad (10)$$

We can then project this vector onto the full family of $\underline{w}$ vectors and obtain a variable measure of the scattered power D, as shown in (10). This optimization problem of max(D) has been solved in [4] and here we show the form of the solution (the optimum weight vector $\underline{w}$ to use) in (11).

$$\begin{cases} z_2 = \underline{e_2}^{*T}\underline{k} \\ z_3 = \underline{e_3}^{*T}\underline{k} \end{cases} \rightarrow \begin{cases} \tau = arg(z_2^* z_3) \\ tan2\gamma = \frac{2|z_2||z_3|}{|z_2|^2-|z_3|^2} \end{cases} \rightarrow \underline{w}_{opt} \quad (11)$$

Here we first project the sample $\underline{k}$ vector onto the reference states $\underline{e_2}$ and $\underline{e_3}$ and then generate $\underline{w}$, as a weighted sum of vectors using the spherical angles $\gamma$ and $\tau$ as shown. This then yields the maximum signal $D_{max}$. To illustrate, consider a numerical example. In (12) we show a normalized sample Pauli vector for a selected image pixel. For this we want to maximise $D(\underline{w})$ in the null space of mirror reflections (see (6)). This fixes $\underline{e_2}$ and $\underline{e_3}$, but we can then move over the ortho-sphere to see how the residual signal D varies.

$$\underline{k} = \begin{bmatrix} 0.3447 \\ 0.6650 + 0.4036i \\ 0.1579 - 0.5011i \end{bmatrix} \rightarrow |\underline{k}| = 1 \quad (12)$$

Fig. 4 shows the variation. Here we see significant changes as we move across the sphere, with copolarized circular shown now as sub-optimum (even though we use a mirror canceller). In fact, the true maximum signal is found using the algorithm in (11) (shown by the black point in fig.4). This still makes the mirror components null, but maximizes the non-mirror components of $\underline{k}$. It is this ability to maximize components seen beneath a background that we seek to exploit.

### A. The General Bistatic Canceller

We can extend the above ideas to higher dimensional vectors [4,5].

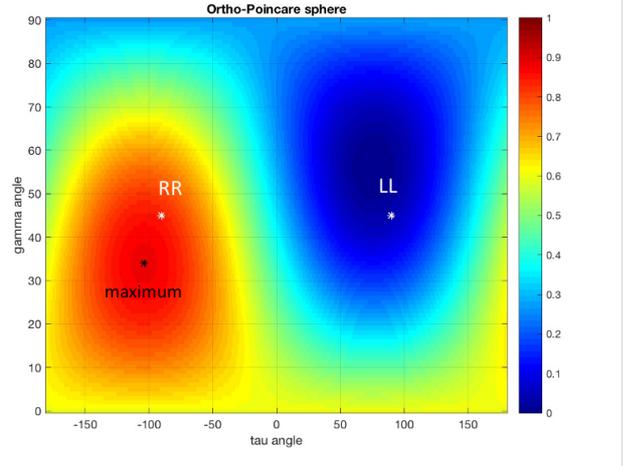

Fig. 4. Variation of scattered power for the numerical example in (12) using the mirror cancellation states of (6). The maximum scattered signal is shown in black, calculated from (11), and importantly is not copol circular.

In particular, if we consider bistatic radar, with separated transmitter and receiver, then S is no longer symmetric and we have an extra channel of information, the HV-VH difference [1]. Here we solve the optimization problem for this more general case to show the potential for increased information extraction in future bistatic POLSAR systems. In this case we now have a 3-dimensional null space [4,5], no longer constrained to the surface of a 3-sphere, but by using complex unitary weights we can still seek a maximum sub-space signal. In (13) we show how to proceed. Now we have 3 basis vectors $\underline{e_2},\underline{e_3}$ and $\underline{e_4}$ and we again project an arbitrary $\underline{k}$ vector onto these states as shown.

$$\underline{k} = \frac{1}{\sqrt{2}}\begin{bmatrix} S_{HH} + S_{VV} \\ S_{HH} - S_{VV} \\ S_{HV} + S_{VH} \\ S_{HV} - S_{VH} \end{bmatrix} \rightarrow \begin{cases} z_2 = \underline{e_2}^{*T}\underline{k} \\ z_3 = \underline{e_3}^{*T}\underline{k} \\ z_4 = \underline{e_4}^{*T}\underline{k} \end{cases} \quad (13)$$

To find the optimum, we set up a 3-compnent unitary weight vector characterized now by 4 angles, as shown in (14).

$$\underline{w}_{opt} = cos\gamma \underline{e_2} + sin\gamma cos\chi e^{i\tau_3}\underline{e_3} + sin\gamma sin\chi e^{i\tau_4}\underline{e_4} \quad (14)$$

The optimum weights (to secure $D_{max}(\underline{w})$) can then be found using the algorithm shown in (15) [4].

$$\begin{cases} \tau_3 = arg(z_3 z_2^*), \tau_4 = arg(z_4 z_2^*) \\ tan2\chi = \frac{2|z_3||z_4|}{|z_3|^2-|z_4|^2} \\ tan2\gamma = \frac{2(cos\chi|z_2||z_3|+sin\chi|z_2||z_4|)}{(|z_2|^2-cos^2\chi|z_3|^2-sin^2\chi|z_4|^2-sin2\chi|z_3||z_4|)} \end{cases} \quad (15)$$

There are two key observations to be made for bistatic systems. First is that now we have a much larger space to search for optimum signals, 3-dimensional instead of 2-dimensional complex space, and hence we can exploit many more types of scattering for applications. Secondly however, we can also consider the idea of using rank-2 reference coherency matrices to be nulled, rather than rank-1. In other words, we can include some depolarization properties of a reference target and still remain orthogonal, albeit now again in a 2-D search space of the ortho-sphere in fig. 3. Note that this is not possible for monostatic systems, where, if we wish to consider rank-2 reference states, there is only ever 1 unique orthogonal state and so no space in which to optimize. This suggests a rich vein of new possibilities for future bistatic POLSAR systems. For monostatic we must retain the rank-1 reference and now turn to consider how to generate such reference states from SLC POLSAR monostatic data.

## IV. RANK-1 POLSAR PROCESSING

POLSAR data is often provided in single-look complex (SLC) format, when each pixel has a natural rank-1 form (i.e. a single S matrix). However, for low to medium resolution systems, speckle means that such S matrices show large variance from pixel to pixel and so some form of speckle filtering is generally employed via local averaging. This leads to estimation of a local coherency matrix T, which is no longer rank-1 and so as a first step we need to restore the rank-1 approximation to enable orthogonal processing. This we can do in several ways, but here we employ an eigenvalue decomposition of each T, as shown in (16) [1].

$$\bar{T} = \sum_{j=1}^{3} \lambda_j \underline{e}_j \underline{e}_j^{*T} \rightarrow \lambda_1 \geq \lambda_2 \geq \lambda_3 \geq 0 \quad (16)$$

For each (averaged) pixel we then select only the dominant or maximum eigenvalue and its associated eigenvector $\underline{e}_1$, as shown in (17) [4].

$$T_1 = \lambda_1 \underline{e}_1 \underline{e}_1^{*T} \quad (17)$$

If we wish to cancel at the pixel level, then this is sufficient, the null space fully defined by $\underline{e}_2$ and $\underline{e}_3$. However, often we wish to cancel not a single pixel but a whole segment, comprising a number of pixels in the image, related to a target of interest, such as fields of a particular crop for example. Fig. 5 shows an example of segmented land-use types, each containing a large number of speckle-filtered image pixels.

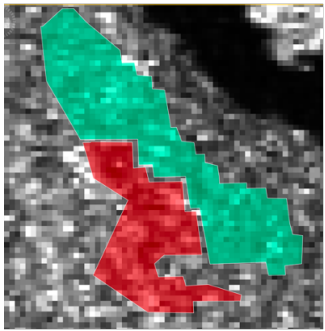

Fig. 5. Sample segments for use as reference null targets enabling orthogonal processing for land-use applications

In this case we can find the reference null states by first forming the average segment coherency matrix $T_{seg}$, as shown in (18). Note that this an average of rank-1 matrices, not the average of full coherency matrices for the segment.

$$T_{seg} = \frac{1}{N} \sum_{i=1}^{N} (\lambda_1 \underline{e}_1 \underline{e}_1^{*T})_i \quad (18)$$

We can again use an eigenvalue decomposition of this segment matrix and select the dominant eigenvector as the reference for null space selection, as shown in (19). Note that these eigenvalues $\mu_i$, are not the same as the full coherency matrix values $\lambda_i$ in (16). However, they are useful, as the ratio $10\log_{10}(\mu_2/\mu_1)$ can be used for example to assess the potential intra-class isolation of the target segment (the level of agreement between pixels under the rank-1 approximation). If the segment shows a wide variety of rank-1 pixels, then this ratio will approach 0 dB and the smaller it is, the better for orthogonal processing. This gives the user a useful metric to assess the potential for using orthogonal processing in their application. We can then use the smaller eigenvectors $\underline{v}_2$ and $\underline{v}_3$ of this segment average matrix as a basis for constructing the null space.

$$T_{seg} = \sum_{j=1}^{3} \mu_j \underline{v}_j \underline{v}_j^{*T} \rightarrow \mu_1 \geq \mu_2 \geq \mu_3 \geq 0 \quad (19)$$

Fig. 6 shows application of these ideas to the ALOS-2 scene for NW Scotland, first shown in fig. 3. The data is multi-looked using a 7x5 (azimuth x range) average. Each pixel is then approximately 20m square on the ground. A 3x3 boxcar filter is then used to reduce speckle further.

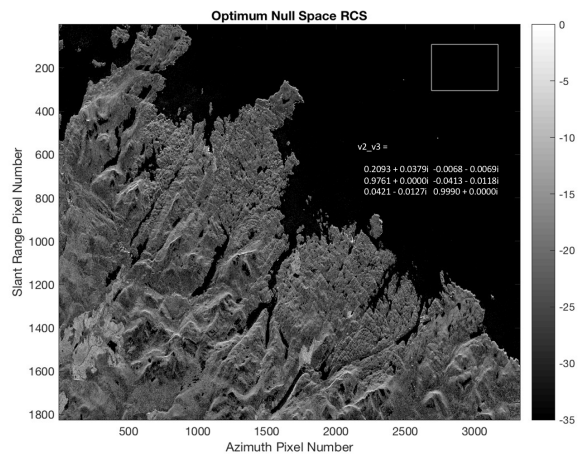

Fig. 6. Optimum null space backscatter image of ALOS-2 scene of NW Scotland (see fig. 3), using pixels inside the ocean segment box shown top right as null reference.

We then select a segment of ocean scattering as a box shown top right. After averaging all rank-1 pixels inside the segment, we fix the basis for a null space as $\underline{v}_2$ and $\underline{v}_3$ as shown in fig. 6. We can also estimate the variation within the box using the μ values. At L-band the ocean scattering is very polarized and so we obtain a small null ratio of -29dB for this segment. This low value provides very good cancellation of the whole ocean scene (and emphasizes the ship targets) as shown in fig.6. Note how much better this null space optimum is at ocean suppression than the simple mirror cancellation of LL in fig. 2. We note also that it has a variable effect on the land surfaces, to which we now turn.

## V. EXAMPLE: POLSAR LAND-USE SEGMENT ISOLATION USING ORTHOGONALITY

We now turn to give an example of application of the null space optimization to land-use studies. We employ again ALOS-2 L-band POLSAR data, this time for a mixed scene around the city of Glasgow in Scotland, collected 7[th] March 2017. We again use a 7x5 multi-look with 3x3 boxcar filtering and rank-1 pre-filtering of each pixel before selecting different land segments. Fig.7(a) shows a 'standard' HV or cross-polarized image of the scene (obtained before rank-1 filtering). We see a mixed scene with urban areas, lakes, semi-natural moorland, forest and agriculture. This provides a good set of land-use classes to illustrate the new processing technique. We now show three processed images of the *same* data set. Each is chosen to null a particular type of land-cover, using different training segments (shown in white boxes). In fig. 7(b) we show cancellation of open moorland and can see strong contrast with urban areas, and also see clearly an enhanced array of point scatterers on the moorland, being wind turbines at Whitelee, the UK's largest on-shore windfarm, with more than 200 turbines visible. In contrast, fig. 7(c) shows cancellation of all urban areas, displaying very different image contrast to fig. 7(b), with suppression of strong urban reflections. Finally, in fig. 7(d) we show that even random scatterers such as forests can be cancelled using this technique. They too have a stable rank-1 approximation at L-band, allowing them to be removed from the SAR imagery, to obtain essentially a forest-free SAR image.

## VI. CONCLUSIONS

In this paper we have developed a new method for processing polarimetric synthetic aperture (POLSAR) data. Here we prioritize the ability to exploit orthogonality in a scene rather than maximise multi-channel data for classification. We have shown how to solve an important optimization problem in the null space of any scatterer, namely, how to maximise residual radar cross section (RCS). We developed analytical solutions for this optimization for both mono and bistatic POLSAR systems. We illustrated the new technique using L-band POLSAR data from the JAXA ALOS-2 system and illustrated not only improved point target detection (ships at sea), but also showed how land-use classes can be used to provide strong contrast in POLSAR data.

In order to help the wider SAR image community explore this new approach, we have developed accompanying software for open distribution. Details can be found here ( https://github.com/ashlinrichardson/cloude_decom).

In this way users can first pre-process any scene of POLSAR data to generate multi-look T3 coherency matrix form, using for example the free ESA-SNAP (https://step.esa.int/main/download/snap-download/)
or Polsarpro (https://step.esa.int/main/toolboxes/polsarpro-v6-0-biomass-edition-toolbox/) software packages and then implement and explore the ideas of generalized Poincaré orthogonality for their own application.

## VII. ACKNOWLEDGMENTS

Thanks to the Japanese Space Agency (JAXA) for provision of the ALOS-2 data used in this project.

## VIII. REFERENCES

[1] S R. Cloude, "Polarisation: Applications in Remote Sensing", Oxford University Press, 978-0-19-956973-1, 2009

[2] I. Hajnsek, Y.L Desnos (Eds), "Polarimetric Synthetic Aperture Radar:Principles and Applications", Springer, open access, 2021

[3] S. R. Cloude, D. Goodenough, H. Chen, "Compact Decomposition Theory", IEEE Geoscience and Remote Sensing Letters, Vol. 9 (1), pp 28-32, Jan 2012

[4] S. R. Cloude, "Target Detection Using Rank-1 Polarimetric Processing," IEEE Geoscience and Remote Sensing Letters, Vol. 18, no. 4, pp. 717-720, April 2021

[5] S R Cloude, "Depolarization Synthesis: understanding the optics of Mueller matrix depolarization", Journal of the Optical Society of America, JOSA A, Vol 30, pp 691-700, April 2013

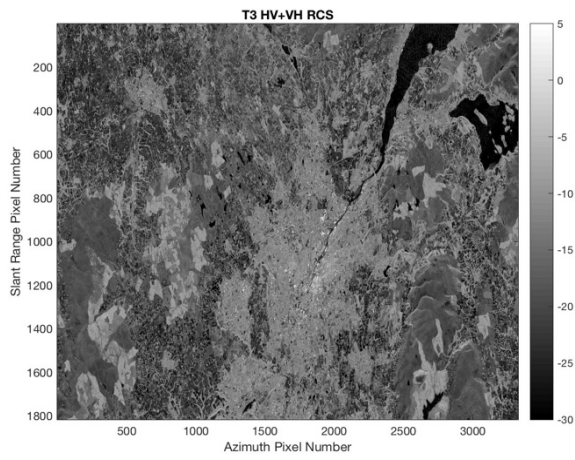
(a)
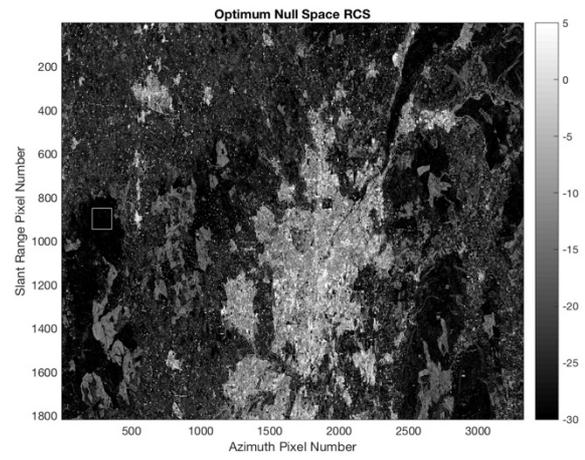
(b)
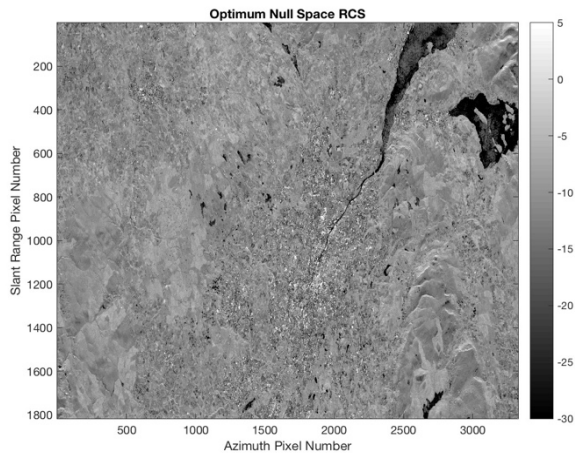
(c)
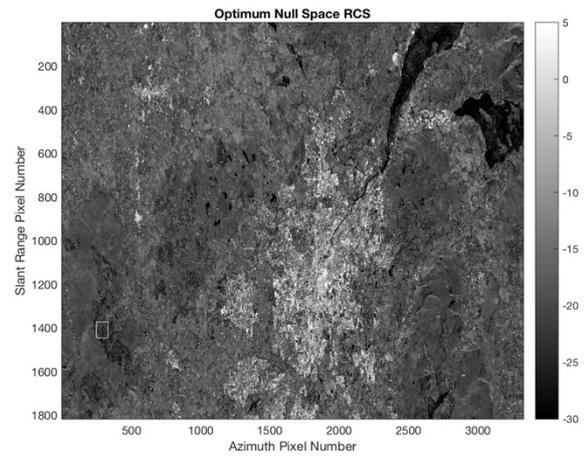
(d)

Fig. 7: JAXA ALOS-2 Glasgow POLSAR scene (collected 7[th] March 2017): (a) is 'standard' full rank HV+VH RCS image for region around city of Glasgow, Scotland showing diversity of land-cover types. (b) optimum image after cancellation of moorland features, (c) optimum image after cancellation of urban areas and (d) optimum image after forest cancellation